\documentclass[11pt]{article}
\usepackage[utf8]{inputenc}
\usepackage[T1]{fontenc}  
\usepackage[colorlinks=true, citecolor=calpolypomonagreen]{hyperref}
\usepackage{url} 
\usepackage{microtype}  
\usepackage[top=2.6cm, bottom=2.6cm, left=1.9cm, right=1.9cm]{geometry} 
\usepackage{tikz}
\usepackage{caption}
\usepackage{algorithm}
\usepackage{algorithmic}
\usepackage{minibox}
\usepackage{mdframed}
\usepackage[sc]{mathpazo}
\usepackage{xcolor}   
\definecolor{calpolypomonagreen}{rgb}{0.12, 0.3, 0.17}
\definecolor{ballblue}{rgb}{0.36,0.54,0.66}
\usepackage{graphicx}    
\usepackage{multicol}
\usepackage{float}
\usepackage{dirtytalk}
\usepackage{enumitem}
\usepackage{mathtools}
\usepackage{amsmath}
\usepackage{amsthm}
\usepackage{subfigure}
\usepackage{physics}
\usepackage[classfont=bold,langfont=caps,funcfont=roman]{complexity}
\usepackage[libertine]{newtxmath}        
\theoremstyle{definition}
\newtheorem{definition}{Definition}[section]

\theoremstyle{plain}
\newtheorem{theorem}[definition]{Theorem}
\newtheorem{lemma}[definition]{Lemma}
\newtheorem{corollary}[definition]{Corollary}

\newtheorem*{theorem*}{Theorem}
\newtheorem*{corollary*}{Corollary}
\theoremstyle{remark}

\usepackage{cleveref}
\usepackage[style=ieee-alphabetic]{biblatex}
\newcommand{\cH}{H}

\newcommand{\email}[1]{\href{mailto:#1}{\textsf{#1}}}

\title{Property Testing in Bounded Degree Hypergraphs}
\date{Department of Computer Science and Technology, University of Cambridge
\\ Cambridge, United Kingdom \vspace{0.8cm}\\
\today}

\author{\textbf{Hugo Aaronson} \footnote{Corresponding author. Email: \email{ha406@cam.ac.uk}} \hspace{0.6cm}
\textbf{Gaia Carenini} \footnote{Corresponding author. Email: \email{gc645@cam.ac.uk}}\hspace{0.6cm}
\textbf{Atreyi Chanda}} 

\begin{document}
\maketitle

\begin{abstract}
We extend the bounded degree graph model for property testing introduced by Goldreich and Ron (\emph{Algorithmica}, 2002) to hypergraphs. In this framework, we analyse the query complexity of three fundamental hypergraph properties: colorability, $k$-partiteness, and independence number. We present a randomized algorithm for testing $k$-partiteness within families of $k$-uniform $n$-vertex hypergraphs of bounded treewidth whose query complexity does not depend on $n$. In addition, we prove optimal lower bounds of $\Omega(n)$ on the query complexity of testing algorithms for $k$-colorability, $k$-partiteness, and independence number in $k$-uniform $n$-vertex hypergraphs of bounded degree. For each of these properties, we consider the problem of explicitly constructing $k$-uniform hypergraphs of bounded degree that differ in $\Theta(n)$ hyperedges from any hypergraph satisfying the property, but where violations of the latter cannot be detected in any neighborhood of $o(n)$ vertices. 
\end{abstract}

\newpage

\section{Introduction}
Many algorithmic settings prioritise performance measures like \emph{space efficiency} or \emph{time complexity} when looking for exact solutions to decision problems. However, when we work with increasingly large objects, it can be impossible to process the full input. This motivates the study of \emph{sublinear algorithms}, i.e.,  algorithms that make decisions based on only small local portions of the input. 


A well-studied subfield of sublinear algorithms is \emph{property testing}, the area that addresses the issue of determining with high probability whether an object has a pre-determined property or is "far" from any object having it. The idea first appeared implicitly in the work of \citeauthor{Blum1993} \cite{Blum1993}, which introduced a tester for group homomorphisms. Further developments in this direction can be found in \cite{Gemmell1991, RS1992}, with \citeauthor{RS1996} \cite{RS1996} making notable progress by partially abstracting the approach. However, a more systematic study of property testing began with \citeauthor{goldreich1998property} \cite{goldreich1998property}, who helped establish it as a distinct class of computational problems rather than just a tool for program checking or PCP system development (see \cite{goldreich2010introduction} for a historical overview). 

 
 Since its introduction, property testing has led to the development of algorithms across a wide range of settings. A significant focus of research, pioneered by \citeauthor{goldreich1998property} \cite{goldreich1998property}, has been on combinatorial structures, particularly graphs (see \cite{NCL2011} for a survey). In \emph{graph property testing}, given \emph{oracle-access} to a graph $G$, one must determine whether $G$ satisfies a property $\mathcal{P}$ or is $\varepsilon$-far from satisfying it. This means that at least an $\varepsilon$-fraction of $G$ must be "modified" in order to obtain a graph satisfying $\mathcal{P}$. The definition of "modified" here depends on the underlying model which represents this graph. Two models -- corresponding to two alternative representations -- are usually taken into consideration in graph property testing: the \emph{dense model}, and the \emph{bounded degree model} introduced by \citeauthor{goldreich1998property} \cite{Gol1997, goldreich1998property}. In the dense model, a graph is represented by its adjacency matrix. In contrast, the bounded degree model represents graphs by their adjacency lists of fixed length. Query access is given to the entries of the adjacency matrix or respectively to those of the adjacency lists.

The dense model is well suited for the study of dense graphs; it is a well-established framework where much is already known: importantly, Alon, Fischer, Newman and Shapira obtained a complete characterization of the properties that are testable with a constant number of queries using \emph{Szemeredi's graph regularity lemma} \cite{AFN+2006}. Here, the adjective "constant" stands for independent from the size of the graph. On the other hand, the bounded degree model is a better framework for studying properties of sparse graphs. Yet, this model is significantly less understood. A handful of testing algorithms are known for specific graph properties (see for instance
\cite{GR99,CS2007,yoshida2015testing})
as well as a few lower bounds, such as for 3-\emph{colorability}, or the \emph{Hamiltonian cycle} problem \cite{BOT02, Gol2020}. However, only few results with a more general flavor have been obtained so far, and each of them refers either to a restricted class of properties, or to limited families of graphs (e.g., see \cite{czumaj2007testable, kohler2021logical})
. Unfortunately, a full characterization of properties testable with a constant number of queries seems out of reach for current techniques (see \cite{goldreich2021open} and its references for an insightful discussion of the problem). 

Hypergraphs are a natural generalization of graphs, where edges can contain more than two vertices. Despite their broader applicability, hypergraphs have received significantly less attention in the field of property testing. The systematic study of \emph{hypergraph property testing} for dense hypergraphs was initiated by \citeauthor{CS05}, who  exhibited a tester for \emph{hypergraph colorability} \cite{CS05}.\footnote{Previously, few results were known for 3-uniform hypergraphs \cite{KNR2002,ARS2005}.} Soon after this work, \citeauthor{RS2007} proved that all hereditary graph properties are strongly testable, and more recently \citeauthor{Joos2017} fully characterized hypergraph properties efficiently testable in the dense model \cite{Joos2017}, generalizing the results in \cite{AFN+2006}.  Lastly, we remark that \citeauthor{Espuny2019} -- while studying the problem of testing the containment of copies of fixed subgraphs in hypergraphs --introduced a testing framework that generalizes graph general model \cite{Espuny2019}. 

As far as we are aware, no prior work has explicitly formalized property testing for hypergraphs of bounded degree. This generalization is natural since sparse hypergraphs are common in a wide variety of applications, and since the bounded degree model was previously successful when used for testing objects closely related to hypergraphs, such as CSPs \cite{Yos2011}. In this work, we address this gap in the literature by introducing the bounded-degree model for hypergraph property testing, and proving the first upper and lower bounds in this framework.

\subsection{Contributions}

For a set $V$ and an integer $k\geq 1$, let $\binom{V}{k}$ be the set of all $k$-element subsets of $V$. We call $H=(V(H), E(H))$, where $E(H)\subseteq \bigcup_{k=1}^{|V(H)|}\binom{V(H)}{k}$, a \emph{hypergraph} over the set $V(H)$. We denote by $V(\cH)$ and $E(\cH)$ its \emph{vertex} and \emph{hyperedge set}, respectively. In this work, we consider only \emph{finite} hypergraphs (i.e., $V(H)$ is finite) and we may assume that all $n$-vertex hypergraphs have $V(H)=\{1,\dots, n\}$. If the hyperedge set $E(H)$ contains exclusively sets of size $k$, we say that $H$ is a $k$-uniform hypergraph. The \emph{degree} of a vertex $v$ is defined as the number of hyperedges incident on $v$. We denote by $\Delta(\cH)$ the maximum degree in $\cH$.

In this work, we focus on $k$-uniform hypergraphs of \emph{bounded degree}, i.e., we assume that $\Delta(\cH)$ is constant, and we fix an arbitrary yet global ordering $\mathcal{O}$ of the hyperedges in $E(H)$. We represent a hypergraph $H$ with bounded degree $\Delta$ as a list of lists of length $\Delta$, denoted by $A(H)$, and defined as follows $A(H)=(A_{1},\ldots, A_{n})=((A_{1,1},\ldots,A_{1,\Delta}),\ldots, (A_{n,1},\ldots,A_{n,\Delta})),
$ where, for all $i\in [n]$, $A_{i}$ is the list containing the $\text{deg}(i)$ hyperedges $A_{i,1},\dots, A_{i,\text{deg}(i)}$ adjacent to $i$, and $\Delta-\text{deg}(i)$ dummy hyperedges $A_{i,\text{deg}(i)+1},\dots, A_{i,\Delta}$, represented by the symbol $\{\bot\}$. Here, we may assume that the hyperedges are enumerated consistently with the global ordering $\mathcal{O}$.   

We measure the distance between two hypergraphs $H_1 = (V_1,E_1)$ and $H_2 = (V_2,E_2)$ of bounded degree $\Delta$, denoted by $d_H(H_1, H_2)$, is defined as the proportion of hyperedges that need to be added or removed to turn $H_1$ into $H_2$, i.e., 
    \begin{equation*}
        d_H(H_1, H_2)=\frac{|E_1\setminus E_2|+|E_2\setminus E_1|}{\Delta n}.
    \end{equation*}

A \emph{hypergraph property} $\mathcal{P}$ is a predicate over hypergraphs that is preserved under hypergraph isomorphism, meaning that if a hypergraph $H$ has the property $\mathcal{P}$ then any hypergraph that is obtained by relabeling the vertices of $H$ also has $\mathcal{P}$. Let $\varepsilon\in (0,1)$, $k\geq 1$, $\mathcal{H}_{n,k}$ be the family of all $k$-uniform $n$-vertex hypergraphs, and let $\mathcal{H}_{n,k}^{\mathcal{P}}$ be the subset of $\mathcal{H}_{n,k}$ consisting of all hypergraphs satisfying $\mathcal{P}$. We say that a hypergraph $H$ is $\varepsilon$-\emph{far} from a property $\mathcal{P}$ if the following holds:
\begin{equation*}
d_H(H,\mathcal{P}):=\min_{H'\in \mathcal{H}_{n,k}^{\mathcal{P}}}d_H(H,H')>\varepsilon   
\end{equation*}

For a fixed degree bound $\Delta$, a $\varepsilon$-\emph{tester} for a hypergraph property $\mathcal{P}$ is a probabilistic oracle machine that, on input parameters $n$ and $\varepsilon$, and oracle access to a hypergraph $H=([n], E)$ of maximum degree $\Delta$, outputs a binary verdict that satisfies the following two conditions:
\begin{itemize}
    \item[(i)] If $H\in \mathcal{H}_{n,k}^{\mathcal{P}}$, then the tester accepts with probability at least 2/3; 
    \item[(ii)] If $H$ is $\varepsilon$-far from having property $\mathcal{P}$, then the tester accepts with probability at most 1/3. 
\end{itemize}
We say that a tester has 1-\emph{sided error} if it always accepts when $H$ has $\mathcal{P}$, and 2-\emph{sided error} otherwise. Note that the behavior of the tester might be arbitrary in the regime in which $H$ neither has property $\mathcal{P}$ nor is $\varepsilon$-far from having the property. 

The complexity measure we consider for testers is its \emph{query complexity}, defined as the function of the parameters $\Delta,n$, and $\varepsilon$ that represents the number of queries made by the tester to the oracle on the worst-case $n$-vertex hypergraph of maximum degree $\Delta$, once fixed the proximity parameter $\varepsilon$. Fixing $\Delta$, we may treat it as a hidden constant. We say that a hypergraph property $\mathcal{P}$ is \emph{strongly testable}, if there exists a tester for $\mathcal{P}$ such that its query complexity is independent from the size of the hypergraph, and that the query complexity of testing $\mathcal{P}$ is $\Omega(n)$, when there exists a constant $\varepsilon>0$ such that distinguishing between $n$-vertex hypergraphs in $\mathcal{P}$, and $n$-vertex hypergraphs that are $\varepsilon$-far from $\mathcal{P}$ requires $\Omega(n)$ queries.

In this paper, we analyse the query complexity of three fundamental hypergraph properties: \emph{colorability}, $k$-\emph{partiteness}, and \emph{independence number}\footnote{Note that all our hardness results can be easily extended to two-sided error testers, while our upper bound holds for one-sided error testers. This is the case since our lower bounds for the query complexity of testers are given via gap-preserving reductions to graph problems for which lower bounds for two-sided error testers are known. This is discussed in Section \ref{2-sided-exp}.} 

Given a set of colors $\{1,2,\dots, \lambda\}$, we say that the hypergraph $\cH$ is  $\lambda$-\emph{colorable} if there exists a map $\mathcal{C}:V(\cH)\rightarrow\{1,2,\dots, \lambda\}$ that associates a color to every vertex of a hypergraph such that every hyperedge contains at least two vertices with different colors, and that $\cH$ is  $k$-\emph{partite} if there exists a map $\mathcal{C}_r:V(\cH)\rightarrow\{1,2,\dots, k\}$ that associates a color to every vertex of a hypergraph such that every hyperedge contains exactly $k$ vertices with distinct colors. A set of vertices is an \emph{independent set} if no hyperedge is fully contained within it. We define the \emph{independence number} of a hypergraph $H$ as the size of a maximum independent set in $H$. 

First, we show that $k$-coloring $k$-uniform hypergraphs is maximally hard, i.e., we prove that its query complexity is linear in $n$, the number of vertices of the hypergraph. We achieve this by reducing this problem to the one of testing the satisfiability 
of $(3, d)$-SAT, a 3-SAT instance where each literal appears in at most $d$ clauses. This reduction is novel, and can be interpreted as a (non trivial) generalization to hypergraphs of the reduction previously employed by \citeauthor{BOT02} to show the hardness of testing graph 3-colorability in the bounded degree model \cite{BOT02}. Specifically, we prove the following.

\begin{theorem}[Hardness of testing hypergraph $k$-colorability]\label{theo: Hardness-k-col}
For every $k\geq 3$, there are positive constants $d_{k\text{-col.}}, \varepsilon_{k\text{-col.}}$, such that the query complexity of testing $k$-colorability of $k$-uniform $n$-vertex hypergraphs of bounded degree $d_{k\text{-col.}}$ is $\Omega(n)$.
\end{theorem}

Observe that Theorem \ref{theo: Hardness-k-col}, combined with earlier work by \citeauthor{CS05} \cite{CS05} showing that colorability of dense hypergraphs is testable (with one-sided error) using $\poly(1/\varepsilon)$ queries underscores a very strong separation between the dense and bounded degree models in hypergraph property testing, that reflects the one known for graphs.  Namely, we can deduce the following corollary:

\begin{corollary}[Dense vs. bounded degree model for hypergraphs, informal]
There exists a hypergraph property $\mathcal{P}$ that admits a one-sided $\varepsilon$-tester whose query complexity in the dense model does not depend on $n$, the number of vertices in the hypergraph, while its query complexity in the bounded degree model is $\Omega(n)$. 
\end{corollary}

Second, we address the problem of testing 
hypergraph $k$-partiteness and show that it is also maximally hard. A straightforward reduction from graph coloring achieves this result. As a byproduct of our technique, we manage to prove a slightly stronger result, namely that the query complexity of $k$-partiteness is maximal even when we exclusively consider \emph{simple hypergraphs}, i.e., hypergraphs such that every pair of vertices is included in at most a hyperedge. This result stands in stark contrast to the one known for \emph{graph bipartiteness} in the bounded degree model, that is indeed testable with $\sqrt{n}$ queries, where $n$ is the number of vertices in the graph \cite{GR99}. 

\begin{theorem}[Hardness of testing hypergraph $k$-partiteness]\label{theo: Hardness-k-part}
For every $k\geq 3$, there are positive constants $d_{k\text{-par.}}, \varepsilon_{k\text{-par.}}$, such that the query complexity of testing $k$-partiteness of $k$-uniform $n$-vertex hypergraphs of bounded degree $d_{k\text{-par.}}$ is $\Omega(n)$. 
\end{theorem}

Third, we observe that there are reductions between graph coloring and hypergraph $k$-partiteness in both directions, and use the backward direction, together with the fact that testing graph colorability is easy in \emph{non-expanding graph families}  \cite{CSS2009} to pinpoint a hypergraph family where $k$-partiteness is strongly testable. 

\begin{theorem}[Strong testability of $k$-partiteness in bounded treewidth hypergraph]\label{theo: testability-k-par}
For every $k\geq 3$ and for every constant $d_{k\text{-par.}}^{\text{tw.}}$ and $\varepsilon_{k\text{-par.}}^{\text{tw.}}$, $k$-partiteness is strongly testable within the class of $k$-uniform $n$-vertex hypergraphs with bounded degree $d_{k\text{-par.}}^{\text{tw.}}$ and bounded treewidth. 
    
\end{theorem}
Moreover, we consider the problem of testing hypergraph independence number and prove that it requires the maximum number of queries as well. We prove this result by reducing this problem to the one of testing hypergraph $k$-partiteness. This result generalizes the one recently obtained by \citeauthor{Gol2020} for bounded degree graphs \cite{Gol2020}, and contrasts with the one for dense hypergraphs by \citeauthor{Lan2004} \cite{Lan2004}. 

\begin{theorem}[Hardness of testing independence number]\label{theo: ind-num}
For every $k\geq 3$, there are positive constants $d_{\text{ind.}}, \varepsilon_{\text{ind.}}$, $\alpha_{\text{ind.}}$ such that the query complexity of testing whether the independence number of a $k$-uniform $n$-vertex hypergraph is at least $n/\alpha_{\text{ind.}}$ is $\Omega(n)$. 
\end{theorem}

Lastly, as an independently interesting combinatorial question, we consider the problem of constructing hypergraphs that are simultaneously far from satisfying the aforementioned properties, and contains no small subgraphs violating the property. To this end, we adopt the approach suggested by \citeauthor{BOT02} that consists in the use of approximation-preserving reduction in the explicit constructions of a combinatorial objects. In particular, we apply the same reductions we employ to exhibit our lower bounds to explicit construction of suitable instances of $(3, d)$-SAT and graph 3-colorability that were previously presented in \cite{BOT02}.

\subsection{Techniques}

Our proofs follow the general strategy introduced by \citeauthor{BOT02} \cite{BOT02}, and employed by \citeauthor{Gol2020} \cite{Gol2020}. Namely, we apply \emph{polynomial-time reductions} similar to those frequently used to show that a given set is $\NP$-complete, and prove that these reductions are \emph{local} and \emph{gap-preserving}. This means that only objects with a bounded maximum degree are produced via the reduction; few queries to the original instance are needed to answer each incidence query to the resulting object; objects that satisfy the original property are mapped to objects that satisfy the target property; and, objects that are "far" from satisfying the original property are mapped to objects that are also "far" from satisfying the target property. We invite the reader to look at \cite{Gol2020} for a comprehensive explanation of this lower bound technique.

While almost all our reductions are standard, the one from $(3,d)$-SAT to hypergraph $k$-colorability is new since existing reductions weren't local. To circumvent this issue, we rely on suitable expander graphs that avoids degree blow-up. The overall structure of the reduction is inspired by the one employed by \citeauthor{BOT02} to prove the hardness of testing 3-colorability in bounded-degree graphs.

From a technical perspective, one of our contributions consists of showing that the relationships between hypergraph and graph properties can lead to a significant simplification of the analysis of the testability of hypergraph properties. While this approach is natural in graph theory, to the best of our knowledge it hasn't been applied so far in property testing.

\subsection{Structure of the Paper} 
The paper is structured as follows. Sections \ref{sec:colorability}, \ref{sec:partiteness}, and \ref{sec:independence-number} present results for hypergraph $k$-colorability, $k$-partiteness, and independence number, respectively. Lastly, Section \ref{2-sided-exp} describes how to extend our lower bounds to the two-sided error testers and our explicit constructions. While we assume basic familiarity with standard notions of algorithms and computational complexity theory, we review the main concepts needed from graph theory, and property testing in Appendix \ref{sec:preliminaries}. 

\section{Hardness of Testing Hypergraph Colorability}\label{sec:colorability}

This section is devoted to proving that testing $3$-colorability of 3-uniform hypergraphs of bounded degree is maximally hard (Theorem \ref{theo: Hardness-3-col}). Deducing  Theorem \ref{theo: Hardness-k-col} from Theorem \ref{theo: Hardness-3-col} is fairly straightforward as we outline in Appendix \ref{app:kcol}. As \citeauthor{BOT02} mentioned in \cite{BOT02}, Theorem \ref{theo: Hardness-3-col} can be proven via a probabilistic argument. While we work out the details of such a proof in Appendix \ref{app: Probabilistic}, here, we show this fact constructively. The advantage of our proof is that it can be used to get the aforementioned explicit combinatorial constructions. 

\begin{theorem}[Hardness of testing hypergraph $3$-colorability]\label{theo: Hardness-3-col}
For every $n\in\mathbb{N}$, there are positive constants $d_{3\text{-col.}}, \varepsilon_{3\text{-col.}}$, such that the query complexity of testing $3$-colorability of $3$-uniform $n$-vertex hypergraphs of bounded degree $d_{3\text{-col.}}$ is $\Omega(n)$.
\end{theorem}

To prove Theorem \ref{theo: Hardness-3-col}, we rely on a result from \cite{BOT02}. Before stating it, we recall a few basic definitions. Given a Boolean variable $x$ ranging over $\{0,1\}$, its \textit{positive literal} is $x$ and its \textit{negative literal} is $\overline{x}$. A \textit{clause} is a disjunction of literals $\ell_1\vee\dots\vee\ell_t$. Clauses are often viewed as sets: 
the order of the literals is irrelevant, and we can assume there are no repetitions. A  CNF \textit{formula} $\{C_1,\dots,C_m\}$ is a  conjunction of clauses $C_1\wedge\dots\wedge C_m$. A $k$-CNF is a CNF where every clause has at most $k$ literals. We say that a variable $x$ \textit{appears} in a clause $C$ if a literal over $x$ is an element of $C$. A $(k,c)$-CNF is a $k$-CNF formula such that each literal appears in precisely $c$ distinct clauses. A formula is called \textit{satisfiable} if there exists an assignment to the Boolean variables that sets its value to 1, otherwise it is said to be \textit{unsatisfiable}. For a constant $\varepsilon\in(0,1)$, we say that a CNF formula $\rho$ on $m$ clauses is $\varepsilon$-\emph{far} from being satisfiable if every satisfiable CNF formula differs from $\rho$ in at least $\varepsilon m$ clauses.

\begin{theorem}[Hardness of testing $(3,d_{\text{SAT}})$-SAT \cite{BOT02}]\label{theo: BOT-SAT}
There exist positive constants $d_{\text{SAT}}, \varepsilon_{\text{SAT}}$ such that the query complexity of testing satisfiability of $(3,d_{\text{SAT}})$-CNF formulas over $n$ variables is $\Omega(n)$. 
\end{theorem}

The proof of Theorem \ref{theo: Hardness-3-col} is a novel, local, and gap-preserving reduction from $(3,d_{\text{SAT}})$-SAT to 3-colorability in hypergraphs with degree bound $d_{\text{3-col.}}:=d_{\text{3-col.}}(d_{\text{SAT}})$.

Before presenting our argument, we recall the combinatorial definition of \emph{expander graphs}.
Let $G=(V,E)$ be a graph. For each vertex $v\in V$, we denote by $\mathcal{N}(v)$ its \emph{neighborhood}, i.e., the set $\{u\in V:\{u,v\}\in E\}$. We generalize this notion to sets by defining $\mathcal{N}(S)=\bigcup_{v\in S}\mathcal{N}(v)\setminus S$. 
Given a graph $G=(V,E)$ we say that $G$ is a $(n,d)$-expander graph if it is $d$-regular where every subset of vertices $S\subseteq V$ of size at most $|V|/2$, is such that $\mathcal{N}(S)\geq |S|$. It is easy to see via a standard union bound argument that these objects are abundant. Indeed, a $d$-regular random graph is with very high probability an expander graph. While these combinatorial objects have a plethora of applications, here we use them to avoid degree blow-ups in the reduction. 

We now describe an algorithm $\rho_{\text{3-col.}}$ for transforming a $(3,d_{\text{SAT}})$-CNF formula into a $3$-uniform hypergraph of bounded degree $d_{\text{3-col.}}:=d_{\text{3-col.}}(d_{\text{SAT}})$. Then, in Lemma \ref{lem; red-3-col}, we prove that the algorithm $\rho_{\text{3-col.}}$ is a local and gap-preserving reduction. 

\subsection{Description of the algorithm $\rho_{\text{3-col.}}$}
The algorithm $\rho_{\text{3-col.}}$ takes as input a 3-CNF formula, $\mathcal{C}=C_1\wedge\dots\wedge C_m$, over the set of variables, $x_{1},\ldots, x_{n}$, where each literal is contained in $d_{\text{SAT}}$ out of the $m$ clauses, and outputs a 3-uniform hypergraph $H:=\rho_{\text{3-col.}}(\mathcal{C})$. 

 \subsubsection{Overview}
Informally, a hypergraph $H$ must encode all the axioms that hold for satisfiable CNF formulas; namely, that literals have \emph{truth value} either \textsc{True} or \textsc{False}, that every copy of the same literal assumes the same truth value, that for every variable $x$, the positive and the negative literals $x, \neg x$ have opposite truth values, and that for every clause $C=(x\vee y\vee z)$, the literals $x,y,z$ cannot all be simultaneously false. Moreover, any valid truth assignment for the CNF formula $\mathcal{C}$ should be mapped to a coloring of the hypergraph $H$, where the vertices corresponding to the image of literals are colored with the truth values themselves, i.e., with \textsc{True} or \textsc{False}. The coloring is completed by a third color, called \textsc{Dummy}.

\subsubsection{Construction}
We define $V(H)$ as the union of two families of vertices: the \emph{literal vertices}, and \emph{auxiliary vertices}. 

 Namely, there are $4{d}_{\text{SAT}}n$ literal vertices that correspond to $2{d}_{\text{SAT}}$ copies of each of the literals $\mathcal{X}:=\{x_{1},\ldots, x_{n}, \overline{x_1}, \dots, \overline{x_n}\}$ appearing in $\mathcal{C}$, and $24{d}_{\text{SAT}}n$ auxiliary vertices that are partitioned equally into three families corresponding to three distinct colors \textsc{True}, \textsc{False} and \textsc{Dummy}: $T_{\text{aux.}}:=\{T_{1},\ldots, T_{8 {d}_{\text{SAT}}n}\}$, $F_{\text{aux.}}:=\{F_{1},\ldots, F_{8 {d}_{\text{SAT}} n}\}$, and $D_{\text{aux.}}:=\{D_{1},\ldots, D_{8 {d}_{\text{SAT}}n}\}$, respectively. Additionally, we introduce $O(n(d_{\text{SAT}}^{2}+'d_{\text{SAT}}))$ auxiliary vertices $V_{\text{add.}}$ (where $c$ and $c'$ are universal constants) for which we do not fix a color in the construction, and whose colors can be deduced by the constraints.

To define $E(H)$, we have to consider the following families of gadgets:
\begin{itemize}
    \item \emph{\textbf{Equality gadgets}} that ensure that the two distinct vertices $u$ and $v$ are assigned the same color within any valid hypergraph 3-coloring. These gadgets consist of $u,v$, and $5$ auxiliary vertices from $V_{\text{add.}}$, and contain a hyperedge on every set of $3$ vertices that doesn't include both $u$ and $v$. 
    \item \emph{\textbf{Inequality gadgets}} that ensure that two vertices $u$ and $v$ are assigned distinct colors within any valid hypergraph 3-coloring. These gadgets consist of $u,v$ and $12$ auxiliary vertices from $V_{\text{add.}}$, for a fixed pair of auxiliary vertices $u'$ and $v'$, they include equality gadgets between $u$ and $u'$ and $v$ and $v'$, and hyperedges $\{u,u',v\}$ and $\{u,v,v'\}$. 
    \item \emph{\textbf{Not dummy gadgets}} that ensure that a vertex $u$ is not assigned color \textsc{Dummy} in any valid hypergraph $3$-coloring. These gadgets consist of $u$, and two vertices of $D_{\text{aux.}}$, and a unique hyperedge containing them. 
    \item \emph{\textbf{Clause gadgets}} that ensure that any triple of literal vertices $x,y,z$ contained in a clause $C$ is not monochromatically colored with color \textsc{False}. These gadgets consist of two copies of $x,y,z$, as well as $6$ vertices from $T_{\text{aux}}$ and $3$ additional auxiliary vertices from $V_{\text{add.}}$. For each literal, they include a hyperedge that contains its two copies and an additional vertices. Moreover, for each of these auxiliary vertices, there is a hyperedge that contains it and $2$ vertices from $T_{\text{aux}}$. Lastly, there is a hyperedge that contains the three auxiliary vertices in $V_{\text{add.}}$.  
\end{itemize}
\begin{figure}[h]
\centering
\includegraphics[width=0.6\textwidth]{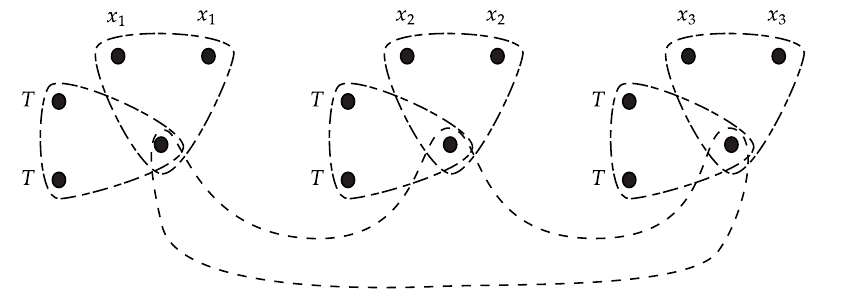}
\caption{Schematic representation of the clause gadget for the clause $(x_1\vee x_2\vee x_3)$.}
\end{figure}

We can now describe how the algorithm $\rho_{\text{3-col.}}$ defines the hyperedges of $H$ via the insertion of the gadgets.  We add an equality gadget between every pair of the $2d_{\text{SAT}}$ copies of each literal. Moreover, we insert equality gadgets among vertices within the same color class $T_{\text{aux.}}, F_{\text{aux.}}, D_{\text{aux.}}$. To do that, we consider an expander graph  $G_{8 d_{\text{SAT}}n}=([8 d_{\text{SAT}} n],E_{8 d_{\text{SAT}} n})$ of degree bounded by a universal constant $s$. For every $(i,j)\in E_{8 d_{\text{SAT}}n}$, we add an equality gadget between $\{T_{i}, T_{j}\},\{F_{i}, F_{j}\},$ and $ \{D_{i}, D_{j}\} $. Then, we insert inequality gadgets among every literal and its negation, in particular, this is done for all copies of the same literal. Moreover, we add inequality gadgets between vertices in distinct color classes $T_{\text{aux.}}, F_{\text{aux.}}, D_{\text{aux.}}$.  As before, we take advantage of $G_{8 d_{\text{SAT}}n}$, and introduce an inequality gadget between $T_{i},F_{i}$ as well as between $F_{i},D_{i}$ and between $D_{i},T_{i}$ for every $i\in [8 d_{\text{SAT}} n]$. Lastly, we add a not dummy gadget for each literal vertex, and for every clause $C\in\mathcal{C}$, we introduce a clause gadget among each triple of literal vertices.

\subsection{Correctness of $\rho_{\text{3-col.}}$}
 We now prove the correctness of the reduction.

\begin{lemma}[Correctness of $\rho_{\text{3-col.}}$]\label{lem; red-3-col}
Algorithm $\rho_{\text{3-col.}}$ defines a local and gap-preserving reduction.
\end{lemma}
\begin{proof}
First, observe that -- by construction -- the number of vertices in the new hypergraph is $O(n)$, and that the degree of every vertex in $H$ is $O(d_{\text{SAT}})$. Therefore, $\rho_{\text{3-col.}}$ always returns a bounded degree hypergraph. Second, note that the locality condition is satisfied. Indeed, one can determine incidences in $H$ by making $O(1)$ queries to $\mathcal{C}$. This is the case since every literal vertex is contained in at most $d_{\text{SAT}}$ clause gadgets, and all other incidences are uniquely defined by $\rho_{\text{3-col.}}$ (i.e., auxiliary vertices aren't used twice in the same family of gadgets). Third, observe that for every satisfiable clause $\mathcal{C}$, the hypergraph $H:=\rho_{\text{3-col.}}(\mathcal{C})$ is $3$-colorable. Specifically, given an assignment $\sigma:\mathcal{X}\to\{$\textsc{True}, \textsc{False}$\}$ satisfying $\mathcal{C}$, we can define a 3-coloring $c:V(H) \to\{$\textsc{Dummy}, \textsc{True}, \textsc{False}$\}$ such that the restriction of $c$ to the literal vertices coincides with $\sigma$, while auxiliary vertices will be assigned the color compatible with all the gadgets. Fourth, note that every $3$-coloring of the hypergraph $H:=\rho_{\text{3-col.}}(\mathcal{C})$, gives a satisfying assignment for $\mathcal{C}$. In particular, let $c': V(H) \to\{$\textsc{Dummy}, \textsc{True}, \textsc{False}$\}$ be a $3$-coloring of $H$, we can obtain a assignment $\sigma':[\mathcal{X}] \to \{$\textsc{True}, \textsc{False}$\}$ satisfying $C$ obtained assigning the literals the color of the corresponding literal vertices.

We now need to strengthen the negative direction of this reduction by showing that $H:=\rho_{\text{3-col.}}(\mathcal{C})$ is far from being 3-colorable whenever $\mathcal{C}$ is far from being satisfiable. To prove that, we can show its contrapositive, i.e., suppose that $H:=\rho_{\text{3-col.}}(\mathcal{C})$ is close to being 3-colorable, then there exists an assignment of $\mathcal{C}$ that has few violations. Let us assume that $H:=\rho_{\text{3-col.}}(\mathcal{C)}$ is $\varepsilon_{\text{3-col.}}$-far from being $3$-colorable, we want to prove that $\mathcal{C}$ is $\varepsilon_{\text{SAT}}$-far from being satisfiable. In order to do that, we can prove the contrapositive namely that if $H:=\rho_{\text{3-col.}}(\mathcal{C)}$ is $(1-\delta_{\text{3-col.}})$-close to $3$-colorability, then $\mathcal{C}$ is $(1-\delta_{\text{SAT}})$-close to satisfiability. 

More formally, suppose that $H$ is $\varepsilon$-close to a 3-colorable hypergraph $H'$, we consider a 3-coloring $c'$ of $H'$. We omit from $H'$ all the hyperedges that are not in $H$, and look at the restriction of the 3-coloring to this subhypergraph. Recall that for an expander graph $G_{t}$ over $t$ vertices, the deletion of at most $\gamma t$ edges with $\gamma<1/2$ leaves a connected component consisting of size $(1-\gamma)t$. We can deduce that deleting $\varepsilon 8d_{\text{SAT}} n$ edges from $G_{8d_{\text{SAT}}}$ leaves each color class with at least $(1-\varepsilon)8d_{\text{SAT}}n$ vertices in a connected component. Then, we still have at least $(1-24\varepsilon)d_{\text{SAT}} n$ triples of vertices $\{D_{i},T_{i},F_{i}\}_{i\in [(1-24\varepsilon)d_{\text{SAT}} n]}$ where for each $i$, all the $D_{i}$s have the same color. The same holds for $T_i$ and $F_i$. Moreover, deleting  $\varepsilon 8d_{\text{SAT}} n$ edges affects at most $48\varepsilon d_{\text{SAT}} n=m(1-48\varepsilon)$ clause gadgets. Since $H'$ is $3$-colorable, then by assigning to the literals of $\mathcal{C}$ the color assigned to the corresponding literal vertices in $H'$, we have that at least a $1-\frac{\varepsilon
}{48}$-fraction of the clauses of $\mathcal{C}$ are satisfied. This concludes the proof of the lemma.
\end{proof}

\section{Testing Hypergraph $k$-Partiteness}\label{sec:partiteness}

The first part of this section is devoted to proving that testing $3$-partiteness of 3-uniform hypergraphs of bounded degree is maximally hard (Theorem \ref{theo: Hardness-3-part}). While the second part of this section is devoted to proving that $3$-partiteness of 3-uniform hypergraphs of bounded degree is strongly testable in bounded treewidth hypergraphs families (Theorem \ref{theo: testability-3-par}). Deducing  Theorem \ref{theo: Hardness-k-part} from Theorem \ref{theo: Hardness-3-part} and Theorem \ref{theo: testability-k-par} from Theorem \ref{theo: testability-3-par} is fairly straightforward as we outline in Appendix \ref{app:kpar}. 

\begin{theorem}[Hardness of testing hypergraph $3$-partiteness]\label{theo: Hardness-3-part}
There exist positive constants $d_{3\text{-par.}}, \varepsilon_{3\text{-par.}}$, such that the query complexity of testing $3$-partiteness of $3$-uniform $n$-vertex hypergraphs of bounded degree $d_{3\text{-par.}}$ is $\Omega(n)$.
\end{theorem}

\begin{theorem}[Strong testability of $3$-partiteness in bounded treewidth hypergraphs]\label{theo: testability-3-par}
For every constant $d_{3\text{-par.}}^{\text{tw.}}$ and $\varepsilon_{3\text{-par.}}^{\text{tw.}}$, $3$-partiteness is strongly testable within the class of $3$-uniform $n$-vertex hypergraphs with bounded degree $d_{3\text{-par.}}^{\text{tw.}}$ and bounded treewidth.
\end{theorem}

\subsection{Hardness of Testing Hypergraph $k$-Partiteness}

The proof of Theorem \ref{theo: Hardness-3-part} is a local and gap-preserving reduction from 3-colorability for graphs of bounded degree $d_{\text{3-col.}}$ to 3-partiteness for hypergraphs of bounded degree $d_{3\text{-par.}}=d_{3\text{-par.}}(d_{\text{3-col.}})$, and relies on the following result by \citeauthor{BOT02}.

\begin{theorem}[Hardness of testing graph $3$-colorability \cite{BOT02}]\label{theo: BOT-Col}
There are positive constants $d_{3\text{-col.}}, \varepsilon_{3\text{-col.}}$, such that the query complexity of testing $3$-colorability of $n$-vertex graphs of bounded degree $d_{3\text{-col.}}$ is $\Omega(n)$.
\end{theorem}

 The reduction is the one classically employed to prove the $\NP$-hardness of the same problem. In particular, we consider the algorithm $\rho_{\text{3-par.}}$ that given a graph $G = ([n], E(G))$ of maximum bounded degree $d_{3\text{-col.}}$ constructs a 3-uniform hypergraph $H:=\rho_{\text{3-par.}}(G)=([n+d_{3\text{-col.}}n], E(H))$ where $E(H)$ is the set of hyperedges $\{u,v, n+(u-1)d_{3\text{-col.}}+j\}$ with $u<v$, and where $(u,v)\in E(G)$ is the $j$th edge incident on $u$ according to an arbitrary yet global ordering of the edges of $G$. 
Lemma \ref{lem; red-3-par} shows that $\rho_{\text{3-par.}}$ is indeed local and gap-preserving. 

\begin{lemma}[Correctness of $\rho_{\text{3-par.}}$]\label{lem; red-3-par}
Algorithm $\rho_{\text{3-par.}}$ defines a local and gap-preserving reduction. 
\end{lemma}
\begin{proof}
First, observe that -- by construction -- the number of vertices in the new hypergraph is $(d_{\text{3-col.}}+1)n$, that the degree of every vertex in $G$ is preserved in $H$, and that every newly introduced vertex has degree at most one. Therefore, $\rho_{\text{3-par.}}$ always returns a bounded degree hypergraph. Second, note that the locality condition is satisfied. Indeed, one can determine incidences in $H$ by making $O(1)$ queries to $G$. In particular, every vertex $n+(u-1)d_{\text{3-col.}}+j$ can only be incident to the hyperedge $\{u,v,n+(u-1)d_{\text{3-col.}}+j\}$ (if $v$ is the j$th$ vertex with an edge incident to $u$ and $u<v$) so you only need to query the $j$th edge incident to $u$ in the original graph. Every other vertex is incident to the image of its incidence list through $\rho_{\text{3-par.}}$. Third, observe that for every 3-colorable graph $G$, the hypergraph $H:=\rho_{\text{3-par.}}(G)$ is $3$-partite. Specifically, given a 3-coloring $c:[n]\to\{1,2,3\}$ of $G$, we can define a 3-partition $p:[n+d_{\text{3-col.}}  n] \to \{1,2,3\}$ such that the restriction of $p$ to the vertices in $[n]$ coincides with $c$, while every vertex $n+(u-1)d_{\text{3-col.}}+j$ will be assigned the color not assigned to $u$ or the $j$th vertex with an edge incident to $u$. Fourth, note that every $3$-partition of the hypergraph $H:=\rho_{\text{3-par.}}(G)$ gives a $3$-coloring for $G$. In particular, let $p: [n+ d_{\text{3-col.}}n] \to \{1,2,3\}$ be a $3$-partition of $H$, we can obtain an admissible $3$-coloring $c:[n] \to \{1,2,3\}$ of $G$ by considering the restriction of $p$ to the vertices in $[n]$. 

We now need to strengthen the negative direction of this reduction by showing that $H:=\rho_{\text{3-par.}}(G)$ is far from being 3-partite whenever $G$ is far from being 3-colorable. To prove that, we can show its contrapositive, i.e., suppose that $H:=\rho_{\text{3-par.}}(G)$ is close to being 3-partite, then there exists a coloring of $G$ that has few violations. More formally, suppose that $H$ is $\varepsilon$-close to a 3-partite hypergraph $H'$, we consider a 3-partition $p'$ of $H'$. We omit from $H'$ all the hyperedges that are not in $H$, and look at the restriction of the 3-partition to this subhypergraph. Since every hyperedge corresponds to a unique edge in $G$, we can define the inverse of $\rho_{\text{3-par.}}$ by taking the first $n$ vertices and for any two vertices, there is an edge between them if there is a hyperedge between them in the hypergraph, we denote this by $\rho_{\text{3-par.}}^{-1}$. Let $G'=\rho_{\text{3-par.}}^{-1}(H')$, this graph is a restriction of $G$ where at most $\varepsilon d_{\text{3-par.}} n$ edges are removed. For every hyperedge $\{u,v,x\}\in H'$, the restriction of $p'$ to $G'$ assigns distinct colors to $u$ and $v$. Therefore, two adjacent vertices in $G'$ receive distinct colors if there is a hyperedge containing $u$ and $v$ in $H'$. Therefore, $G'$ is a $3$-partite graph that is $\varepsilon$-close to $G$, and thus $G$ is $\varepsilon$-close to being $3$-partite. This concludes the proof of the lemma. 
\end{proof}

It is easy to see that hypergraphs obtained from our reduction are always simple. This means that even in this fairly restricted category of hypergraphs testing strong $3$-partiteness is hard.

\subsection{Testability of $k$-Partiteness in Bounded Treewidth Hypergraphs}

Given a hypergraph $H$, its \emph{Gaifman graph} (also called \emph{primal graph} or 2-\emph{section}) $G_H$ is the graph on the vertices of $H$ for which two vertices are connected if and only if they share a hyperedge. Following the literature, we will identify the \emph{treewidth} of a hypergraph $H$ with the one of its Gaifman graph, and therefore we say that a hypergraph $H$ has \emph{bounded treewidth} whenever its Gaifman graph has bounded treewidth (see Appendix \ref{sec:preliminaries} for a formal definition).

The proof of Theorem \ref{theo: testability-3-par} is a local and gap-preserving reduction from $3$-partiteness of hypergraph of bounded degree $d_{3\text{-par.}}^{\text{tw.}}$ and bounded treewidth to $3$-colorability of  graphs of bounded degree $d_{3\text{-col.}}^{\text{tw.}}:=d_{3\text{-col.}}^{\text{tw.}}(d_{3\text{-par.}}^{\text{tw.}})$ and bounded treewidth, and relies on the following result by \citeauthor{CSS2009}. 

\begin{theorem}[Strong testability of $3$-colorability in bounded treewidth graphs \cite{CSS2009}]\label{theo:test-graph}

For every constant $d_{3\text{-col.}}^{\text{tw.}}$ and $\varepsilon_{3\text{-col.}}^{\text{tw.}}$, $3$-colorability is strongly testable within the class of $n$-vertex graphs with bounded degree $d_{3\text{-col.}}^{\text{tw.}}$ and bounded treewidth. 
\end{theorem}
 
 The reduction is the one classically employed to prove the $\NP$-hardness of the same problem. In particular, we consider the algorithm $\rho_{\text{3-par.}}^{tw}$ that given a hypergraph $H=([n], E(H))$ returns its primal graph $G:=\rho_{\text{3-par.}}^{tw}(H)=([n], E(G))$. Lemma \ref{lem; red-3-par-tw} shows that $\rho_{\text{3-par.}}^{tw}$ is indeed local and gap-preserving. 

\begin{lemma}\label{lem; red-3-par-tw}
Algorithm $\rho_{\text{3-par.}}^{tw}$ defines a local and gap-preserving reduction. 
\end{lemma}
\begin{proof}
First, observe that -- by construction -- $|E(G)|\leq3 |E(H)|$, and the degree of every vertex in $G$ is at most twice the degree of the corresponding vertex in $H$. Second, note that the locality condition is satisfied. Indeed, one can determine incidences in $G$ by making $O(1)$ queries to $H$. In particular, the adjacency list of every vertex $u$ in $G$ can be obtained directly from the adjacency list of the same vertex in $H$ by transforming any hyperedge  $\{u,v,z\}$ into the pair of edges $uv$ and $uz$. Third, observe that for every 3-partite hypergraph $H$, the graph $G:=\rho_{\text{3-par.}}^{tw}(H)$ is 3-colorable. Specifically, given a 3-partition $p:[n]\rightarrow\{1,2,3\}$, we define a 3-coloring $c:[n]\rightarrow\{1,2,3\}$ that has as color classes the one defined by the 3-partition, namely $p$ and $c$ are the same mapping. 

We now need to strengthen the negative direction of this reduction by showing that $G:=\rho_{\text{3-par.}}^{tw}(H)$ is far from being 3-colorable whenever $H$ is far from being 3-partite. To prove that, we can show its contrapositive, i.e., suppose that $G:=\rho_{\text{3-par.}}^{tw}(H)$ is close to being 3-colorable, then there exists a 3-partition of $H$ that has few violations. More formally, suppose that $G$ is $\varepsilon$-close to a 3-colorable graph $G'$, we consider a 3-coloring $c'$ of $G'$. We may omit from $G'$ the edges that are not in $G$, and look at the restriction of the 3-coloring to this subgraph. We shall see that this 3-coloring can be used "pretty much as" 3-partition for our hypergraph. Observe that for every hyperedge $\{u,v,z\}\in E(H)$, if there exists a clique among $u,v,z$ in $G'$ (once restricted), then assigning colors given by $c'$ will assure us that $u,v,z$ will be colored with 3 distinct colors. Therefore three vertices  $u, v, z$ contained in a hyperedge do not receive distinct colors only when there is no clique among $u, v, z$ in $G'$. However, since such a clique is present in $G$, and $G$ and $G'$ are $\varepsilon$-close, we can easily deduce that $H$ is also close to being 3-partite. This concludes the proof of the lemma. 
\end{proof}

Note that Theorem \ref{theo:test-graph} is weakening of the statement proven in \cite{CSS2009}. Indeed, \citeauthor{CSS2009}'s result holds for all hereditary properties and does not require the graph family to have bounded treewidth, but rather to be \emph{non-expanding}, i.e., it demands the fact that there exists a constant $n_{\mathcal{F}}$ such that all graphs in the family of size at least $n_{\mathcal{F}}$ are not $(1/ \log^2 n)$-expanders (see Appendix \ref{sec:preliminaries} for a formal definition). As a consequence, our proof entails as well a stronger result. Namely, an equivalent of Theorem \ref{theo: testability-3-par} holds for the family of all hypergraphs whose Gaifman graph is non-expanding; however, since we are not aware of any standard hypergraph family with this property, we decided to state the result in a weaker (yet arguably more natural) form using treewidth. 

\section{Hardness of Testing Hypergraph Independence Number}\label{sec:independence-number}
This section is devoted to proving that testing independence number of 3-uniform hypergraphs of bounded degree is maximally hard (Theorem \ref{theo: ind-num} for $k=3$). Deducing that Theorem \ref{theo: ind-num} holds for all values of $k\geq 3$ is trivial.

Our proof of Theorem  \ref{theo: ind-num} relies on our hardness result for hypergraph $3$-partiteness testing (Theorem \ref{theo: Hardness-3-part}). The proof is a local and gap-preserving reduction from hypergraph $3$-partiteness. In particular, we consider algorithm $\rho_{\text{ind. num.}}$ that given a 3-uniform hypergraph $H = ([n], E(H))$ with bounded degree $d_{3\text{-par.}}$ constructs a hypergraph $H'=([n] \times \{1,2,3\}, E(H'))$, where 
$$E(H'):= \{\{(u,i),(v,i),(w,i)\}\}_{\substack{(u,v,w) \in E\\
i \in  \{1,2,3\}}} \cup \{\{(u,1),(u,2),(u,3)\}\}_{u \in [n]}$$

Lemma \ref{lem: ind. num.} shows that $\rho_{\text{ind. num.}}$ is indeed a a local and gap-preserving reduction.

\begin{lemma}[Correctness of $\rho_{\text{ind. num.}}$]\label{lem: ind. num.}
Algorithm $\rho_{\text{ind. num.}}$ defines a local and gap-preserving reduction. 
\end{lemma}

\begin{proof}
First, observe that -- by construction -- $|E(H')|\leq3 |E(H)|$, and the degree of any vertex in $H':=\rho_{\text{ind. num.}}(H)$ is at most $d_{3\text{-par.}}+1$. Second, note that the locality condition is satisfied. Indeed, one can determine incidences in $H'$ by making $O(1)$ queries to $H$. In particular, the adjacency list of the every vertex $(u,i)$ in $H'$ can be obtained from the adjacency list of $u$ in $H$ by taking the Cartesian product of the vertices in the latter and $\{i\}$, together with the hyperedge $\{(u,1),(u,2),(u,3)\}$. 
Third, observe that for every 3-partite hypergraph $H$, the hypergraph $H':=\rho_{\text{ind. num.}}(H)$ has a independent set of size $n$. Specifically, given a 3-partition $p:[n] \to \{1,2,3\}$, we identify for $i\in [3]$ the set $S_i=\{(v,i): v \in [n] \text{ such that } p(v) = i\}$. Observe that for every pair of vertices $(u,i)$ and $(v,i)$ such that $u\neq v$, there cannot be a hyperedge including both $u$ and $v$ since $p$ is assigning them to the same part. For this reason, $S_i$ is indeed an independent set. 

We now need to strengthen the negative direction by showing that $H':=\rho_{\text{ind. num.}}(H)$ is far from having an independent set of size $n$ whenever $H$ is far from being 3-partite. To prove that, we can show the contrapositive, i.e., suppose that $H':=\rho_{\text{ind. num.}}(H)$ is close to having an independent set of size $n$, then there exists a 3-partition of $H$ with very few violations. More formally, suppose that $H'$ is $\varepsilon$-close to a $H''$ which has an independent set of size $n$, and denote by $S''$ such a independent set. The number of hyperedges determined by vertices in $S''$ in $H'$ is at most $\delta (d+1)n$ for a suitable positive constant $\delta <1$. Note that $ 3 \delta (d+1)n$ upper bounds the number of vertices in $S''$ that can have an incident hyperedge in $H'$.   Let $S'$ be the subset of $S''$ which has no incident hyperedge in $H'$. Clearly, $S'$ is an independent set in $H'$ and $|S'| \geq 1- \varepsilon 3(d+1)n$. From the construction, we can infer that every vertex in $H$ has at most one copy in $S'$. Let $S$ be a subgraph of $H$ induced by every corresponding copy of vertices in $H'$ and let $p$ be a 3-partition of this subgraph such that $p(v) = i$ if and only if $(v,i) \in S'$. Thus, $H$ is $(9(d+1) \varepsilon)$-close to 3-partiteness. This conclude the proof of the lemma. 
\end{proof}

\section{Lower Bounds for Two-Sided Error Testers and Explicit Constructions}\label{2-sided-exp}

Since lower bounds for two-sided error testers are known for $3$-SAT and (graph) $3$-colorability (\cite{BOT02}), we can obtain lower bounds for two-sided error testers for our hypergraph testing problems by applying our reductions to these instances, and if needed by composing these reductions among them. Indeed, it is easy to see that gap-preserving reductions are closed under composition. 

The same procedure allows us to get the explicit constructions. To see that, we recall here the explicit construction of the family of constraint satisfaction problems (CSPs) -- originally introduced in \cite{BOT02} -- that can be used to obtain all the explicit families of graphs and hypergraphs we mentioned. 

We say that a constraint satisfaction problem on $m$ clauses is $(\delta, 1-\varepsilon)$-satisfiable if any subset of at most $\delta m$ constraints is satisfiable, but no assignment satisfies more than $(1-\varepsilon)m$ constraints. For a fixed $d$, consider the $2d$-ary constraints of the form $h:\{0,1\}^d\times \{0,1\}^d\rightarrow \{0,1\}$, where $h(x_1,\dots,x_d,y_1,\dots, y_d)$ is satisfied exactly when $\sum_{i=1}^d x_i=\sum_{i=1}^d y_i+1$, and we identify the Boolean $\{0,1\}$ inputs with integer $0$ and $1$ in a obvious way. We assume to be given an infinite family of $(n,d)$-expanders for some universal constant $d$. We can then define the constraint satisfaction problem $f_n$ on $dn$ variables and $n$ clauses over $h$ as follows. Let $G=(V,E)$ be an $(n,d)$-expander, convert $G$ into a directed multigraph $G'=(V', E')$ by replacing each undirected edge $(i,j)\in E$ with two directed edges $(i,j), (j,i)\in E'$. Each arc $(i,j)\in E'$ is identified with a Boolean variable $x_{i,j}$ in $f_n$. One constraint $h$ is introduced for each $v\in V$, with the predicate variables mapped to the edges incident to $v$:
\begin{equation}
f_n=\bigwedge_{v\in V} h(x_{v, \mathcal{N}^1(v)},\dots, x_{v, \mathcal{N}^d(v)}, x_{\mathcal{N}^1(v),v}, \dots, x_{\mathcal{N}^d(v),v} )
\end{equation}
where for every $i\in [d]$, we denote by $\mathcal{N}^i(v)$ the $i$th neighbor of $v$ according to an arbitrary yet total ordering of the edge set. 
\begin{theorem}[Theorem 16 in \cite{BOT02}]
There exist positive constants $\varepsilon, \delta$ such that the CSP formulas $f_n$ are $(\delta, 1-\varepsilon)$-satisfiable. 
\end{theorem}
Observe that an arbitrary Boolean predicate on a finite number of variables can be expressed as a 3-CNF formula, possibly with introduction of a constant number of auxiliary variables, that the degree of each literal is increased by only a constant factor by this transformation, and thus, there exists a local and gap-preserving reduction from these CSPs that provides us with a family of 3-CNF formula that is also $(\delta, 1-\varepsilon)$-satisfiable. Lastly, note that by applying our reductions in a suitable order together with the reduction from $(3,d)$-SAT to graph 3-colorability, we can obtain the desired explicit constructions.  

\section*{Acknowledgments}
 We thank Tom Gur for bringing our attention to the problem of studying hypergraph property testing in the bounded degree model and for insightful discussion. We also thank Noel Arteche,  Samuel Coulomb, and Ninad Rajgopal for careful proofreading of a draft of this work, and Alberto Espuny Dìaz for suggesting to us important related work. Lastly, we thank the anonymous ECCC board members for identifying issues in the historical account, and helping us addressing them.

The second author is supported by the \emph{CB European PhD Studentship} funded by \emph{Trinity College Cambridge}.
\nocite{GR04}

\printbibliography[heading=bibintoc]
\newpage
\appendix
\section{Background}\label{sec:preliminaries}
We review the definitions from graph theory, and graph property testing omitted in the main body of the paper. 
\subsection{Graph Theory}
For a set $V$, let $\binom{V}{2}$ be the set of all pairs of elements in $V$. We call $G=(V(G),E(G))$, where $E(G)\subseteq\binom{V(G)}{2}$, an (undirected) \emph{graph} over the set $V(G)$. For a given graph $G$, we denote by $V(G)$ and $E(G)$ its \emph{vertex} and \emph{edge set}, respectively.  Given  a vertex $v\in V(G)$, the \emph{degree} of $v$, $\text{deg}(v)$, is defined as the number of edges incident to $v$. We denote by $\Delta(G)$ the maximum degree in the graph $G$. In a \emph{bounded degree graph}, $\Delta(G)$ is assumed to be a constant. 

We define a \textit{clique} in an undirected graph as a subset of vertices pairwise adjacent, meaning that there is an edge between every pair of vertices. On the other hand, an \textit{independent set} is a subset of vertices in which no two vertices are adjacent. 

 For a pair of disjoint vertex sets $V_1$ and $V_2$ in $V$, we denote by $e(V_1, V_2)$ the number of edges connecting vertices from $V_1$ with vertices from $V_2$. For each vertex $v\in V$, we denote by $\mathcal{N}(v)$ its \emph{neighborhood}, i.e., the set $\{u\in V:\{u,v\}\in E\}$. We generalize this notion to sets by defining $\mathcal{N}(S)=\bigcup_{v\in S}\mathcal{N}(v)\setminus S$. A graph $G$ is called $\lambda$-\emph{expander}, if for all $S\subset V$ with $|S|\leq n/2$, we have that $|\mathcal{N}(S)|\geq \lambda |S|$. Unless stated otherwise, we write that a graph is an expander if it is a $\lambda$-expander for $\lambda=1$. We know that families of expanders exist for which every vertex has degree $d$ for some universal constant $d$.

A family of graphs $\mathcal{F}\subseteq S_G$ is called \emph{non-expanding} if there exists a constant $n_{\mathcal{F}}$ such that all graphs in $\mathcal{F}$ of size at least $n_{\mathcal{F}}$ are not $(1/ \log^2 n)$-expanders. In bounded degree graphs, all graph families with good separator properties are non-expanding: indeed, separators and expansion are very intimately connected, and also to the notion of \emph{treewidth}, a well-studied parameter for tractability. We refer to \cite{BPT+2010} for an insightful exposition of these connections. 

In this work, we refer to \emph{bounded treewidth graphs}, therefore -- for the sake of completeness -- we recall the relevant definitions below. We call \emph{tree-decomposition} of $G$ the pair $\mathcal{T}=(T,X)$, where $T=(I,F)$ is a tree and $X=(X_i)_{i\in I}\subseteq\mathcal{P}(V)$ is a family of subsets of $V$, also called \emph{bags}, such that the following properties hold:
\begin{itemize}
    \item [(i)] $\bigcup_{i\in I} X_i=V$;
    \item [(ii)] For all $\{u,v\}\in E$, there exists an $i\in I$ with $\{u,v\}\subseteq X_i$;
    \item [(iii)] For all $v\in V, X^{-1}(v):=\{i\in I|v\in X_i\}$ is connected in $T$. 
\end{itemize}
The \emph{width} of $\mathcal{T}$ is defined as $\text{width}(\mathcal{T}):=\max\{|X_i|:i\in I\}-1$. Then, we define the \emph{treewidth} of the graph $G$ as:
\begin{equation*}
    \text{tw}(G):=\min \{\text{width}(\mathcal{T})|\mathcal{T} \text{tree-decomposition of }G\}
\end{equation*}
We refer the reader to \cite{CFK+2015} for a detailed explaination of this tractability parameter.

Let $\mathcal{G}$ be the set of all possible graphs. We define a \emph{graph property} $\mathcal{P}$ as a subset of $\mathcal{G}$, an example of graph properties is \emph{bipartiteness}, a graph is called \emph{bipartite} if its vertices can be divided into two disjoint sets $U$ and $V$, such every edge connects a vertex in $U$ to one in $V$. We say that a property $\mathcal{P}$ of a graph $G$ is \emph{hereditary}, if it is inherited by the induced subgraphs of $G$, i.e., all graphs formed from a subset of the vertices of $G$, and all of the edges, from the original graph, connecting pairs of vertices in that subset.  A family of graphs $\mathcal{F}$ is called \emph{hereditary} if it is closed under vertex removal. Arguably, the most studied property of graphs is \emph{graph coloring}. Given a set of colors $\{1,2,\dots, \lambda\}$, we say that the graph $G$ is $\lambda$-\emph{colorable} if there exists a map $\mathcal{C}:V(G)\rightarrow\{1,2,\dots, \lambda\}$ that associates a color to every vertex of a graph in such a way that for every edge, the vertices within the edge all have different colors.

\subsection{Property Testing}

In what follows, we assume that a graph $G=(V,E)$ is represented by a set of $|V|$ lists of length $\Delta$. In particular, for every vertex $v\in V$, we are given (as a list) the set of all $\text{deg}(v)$ edges containing $v$, and $\Delta-\text{deg}(v)$ arbitrary symbols $\bot$, that we treat as edges. For two graphs $G_{1}=(V_1, E_1)$ and $G_{2}=(V_2, E_2)$ with degree bound $\Delta$, we define the distance $d_G(G_1, G_2)$ as the proportion of edges that need to be added or removed to turn $G_1$ into $G_2$, i.e., 
    
    \begin{equation*}
        d_G(G_1, G_2)=\frac{|E_1\setminus E_2|+|E_2\setminus E_1|}{\Delta n}
    \end{equation*}

Given a graph property $\mathcal{P}$ and graph $G$, we define the distance between $G$ and $\mathcal{P}$, $d_G(G, \mathcal{P})$ as the minimum distance between $G$ and an element of $\mathcal{P}$, equivalently
    \begin{equation*}
        d_G(G,\mathcal{P})=\min_{G'\in \mathcal{P}}d_G(G,G').
    \end{equation*}

For any constant $\varepsilon\in (0,1)$, we say that a graph $G$ is $\varepsilon$-far from a property $\mathcal{P}$ if $d_G(G,\mathcal{P})>\varepsilon$.

For a fixed degree bound $\Delta$ and a constant $\varepsilon\in (0,1)$, a $\varepsilon$-\emph{tester} for a graph property $\mathcal{P}$ is a probabilistic oracle machine that, on input parameters $n$ and $\varepsilon$, and oracle access to a graph $G=([n], E)$ of maximum degree $\Delta$, outputs a binary verdict that satisfies the following two conditions:
\begin{itemize}
    \item[(i)] If $G\in\mathcal{P}$, then the tester accepts with probability at least 2/3; 
    \item[(ii)] If $G$ is $\varepsilon$-far from $\mathcal{P}$, then the tester accepts with probability at most 1/3. 
\end{itemize}

The \emph{query complexity} of a tester for $\mathcal{P}$ is a function of the parameters $d,n$, and $\varepsilon$ that represents the number of queries made by the tester to the oracle on the worst-case $n$-vertex graph of maximum degree $\Delta$, when given the proximity parameter $\varepsilon$. Fixing $\Delta$, we may treat it as a hidden constant. We say that a graph property $\mathcal{P}$ is \emph{strongly testable}, if there exists a tester for $\mathcal{P}$ such that its query complexity is independent from the size of the graph. We say that the query complexity of a testing problem is $\Omega(n)$, when there exists a constant $\varepsilon>0$ such that distinguishing between $n$-vertex graphs in $\mathcal{P}$ and $n$-vertex graphs that are $\varepsilon$-far from $\mathcal{P}$ requires $\Omega(n)$ queries.

\section{Alternative Lower Bound for Testing 3-Uniform Hypergraph $3$-Colorability}\label{app: Probabilistic}

Let $n\in \mathbb{N}$ be a multiple of 3. We define a distribution $\mathcal{H}$ on $3$-uniform $n$-vertex hypergraphs as follows. First, we sample uniformly at random $d$ partitions of the vertex set $[n]$ into subsets of size three, and denote them by $C_{1},\ldots, C_{d}$. We consider the multi-set union of $C_{i}$, and use it as hyperedge set. For $\{u,v,w)\}\in C_{i}$, we say that $\{v,w\}$ is the $i$th set of neighbors of $u$ in $H$. We denote by $\overline{H}$ the hypergraph obtained by turning every multi-hyperedge in $H$ into a single hyperedge.

In what follows, we prove that with probability $1-o(1)$ a hypergraph $H$ sampled from a specific distribution $\mathcal{H}$ is such that all of its subhypergraphs of size at most $\delta n$ are extremely sparse. Then, we show that extremely sparse hypergraphs are 3-colorable, concluding that each small-enough subgraph of $H$ is 3-colorable. Lastly, we prove that with probability $1-o(1)$ a hypergraph $H$ sampled from distribution $\mathcal{H}$ is far from being $3$-colorable. 

For any set $S\subseteq [n]$, let $H|_S$ the restriction of hypergraphs on $S$. Let $X_S$ be the number of edges in $H|_S$. It is easy to see that the following holds

$$\mathbb{E}[X_{S}]=d{|S| \choose 3}\frac{1}{{n-1 \choose 2}}.$$

\begin{lemma}
    The probability that $H\sim \mathcal{H}$ has an edge with multiplicity greater than $1$ is $o(1)$.
\end{lemma}

\begin{proof}
    For any fixed $i\in [d]$, the probability that any set of the form $\{u,v,w\}\subseteq [n]$ are in partition $C_{i}$ is $1/{n\choose 2}$. Therefore, suppose there is a set $(u,v,w)\in C_{1}$, the probability that it is in another partition is at most $\left(1-\left(1-\frac{1}{{n \choose 2}}\right)^{d-1}\right)=O(\frac{d}{n^2})$. Equivalently, for sufficiently large $n$,
    
    \begin{align*}
        \mathbb{P}[\exists i\in [2,d]:\{u,v,w\}\in C_{i}]\leq \left(1-\left(1-\frac{1}{{n \choose 2}}\right)^{d-1}\right)
        \leq 1-\left(1-\frac{d-1}{{n\choose 2}}\right)
        =\frac{d-1}{{n\choose 2}}.\\
    \end{align*}
Therefore, applying a union bound, we deduce that the total probability that two partitions contain the same hyperedge satisfies the following:
    \begin{equation*}
        \mathbb{P}[\exists \{u,v,w\}\in C_{1}, i\in [2,d]:\{u,v,w\}\in C_{1}\cap C_{i}]=O\left(\frac{d}{n}\right).
    \end{equation*}
So, again by a union bound, we have that the total probability that any element of any partition is also in another is:
    $$\mathbb{P}[\exists i,j\in[d] \text{ with }i\neq j, \{u,v,w\}\subseteq [n]: \{u,v,w\}\in C_{i}\cap C_{j}]=O\left(\frac{d^2}{n}\right)=o(1).$$
This concludes the proof of the lemma. 
\end{proof}

The sparsity condition satisfied by the subhypergraphs of an hypergraph $H$ sampled from $\cH$ was already mentioned in \cite{BOT02}. We recall this fact and provide a simple proof of it.

\begin{lemma}[Lemma 6 in \cite{BOT02}] \label{lem:BOTlem6}

Let $H$ be a hypergraph on $n$ vertices sampled from distribution $\mathcal{H}$. For every constant $K>\frac{1}{2}$, there exists a value $\delta>0$ such that for any set $S\subseteq V(H)$ with $|S|\leq\delta n$ the restriction $H\vert_{S}$ of $H$ have with probability $1-o(1)$ at most $K|S|$ hyperedges.
\end{lemma}
\begin{proof}
First, we assume that the existence of a set $S$ of cardinality $s$ which is including $K s$ (distinct) hyperedges $\{u_{1},v_{1}\, w_{1}\},\dots,\{u_{K  s}, v_{K  s},w_{K  s}\}$. We define $X_{i,k}, Y_{i,k},Z_{i,k}$ as the sets of neighbors of $u_{i},v_{i},$ and $w_{i}$, respectively, in the partition $C_k$. Since for every fixed $q$, sets $X_{p,q},Y_{p,q},Z_{p,q}$ describe the neighbors of at most $6s$ distinct vertices in partition $C_{k}$, we have that for any $i\in [d]$ by a union bound over possible values $k$: 
\begin{equation*}
        \mathbb{P}[\exists k: X_{i,k}\cup Y_{i,k}\cup Z_{i,k}=\{u_{i},v_{i},w_{i}\}|X_{p,q},Y_{p,q},Z_{p,q}:1\leq p\leq i-1,1\leq q\leq d]\leq \frac{d}{(n-6s)^{2}}
\end{equation*}
Therefore: 
    \begin{equation*}
        \mathbb{P}[\forall i:1\leq i\leq d:\exists k:X_{i,k}\cup Y_{i,k}\cup Z_{i,k}=\{u_{i},v_{i},w_{i}\}]\leq \left(\frac{d}{(n-6s)^{2}}\right)^{Ks}\leq \left(\frac{d}{(1-6\delta)^{2}n^{2}}\right)^{Ks}
    \end{equation*}
    For any $s$, the set $S$ can be chosen in ${n\choose s}$ ways, while the set of $Ks$ edges can be chosen in ${{s\choose 3} \choose Ks}$. Therefore, for some constant $s_0$:
    \[
    \begin{split}
        \mathbb{P}[\exists S, s_{0}\leq |S|<\delta   n: |E(G|_{s})|= K  |S|]&\leq \sum_{s=s_{0}}^{\delta   n}{n\choose s}{{s\choose 3} \choose Ks} \left(\frac{d}{(1-6\delta)^{2}n^{2}}\right)^{Ks}\\
        &\leq \sum_{s=s_{0}}^{\delta   n}\left(\frac{ne}{s}\right)^{s}\left(\frac{s^{3}e/6}{Ks}\right)^{Ks} \left(\frac{d}{(1-6\delta)^{2}n^{2}}\right)^{Ks}\\
        &\leq \sum_{s=s_{0}}^{\delta   n}\frac{s^{2Ks-s}e^{Ks+1}d^{Ks}}{n^{2Ks-s}(1-6\delta)^{2Ks}}=o(1)
    \end{split}
    \]
    Additionally, the contribution of set $S$ of size less than $s_{0}$ is also $o(1)$ as $s_{0}$ is a constant.
\end{proof}

Intuitively, very sparse graphs should be easy to color since each vertex has very few neighbors. This intuition can be made quantitative as follows. 
\begin{lemma}
There exists a value $\delta>0$ such that for every set $S$ such that $|S|\leq \delta n$ the restriction $H|_{S}$ of $H\sim \mathcal{H}$ is 3-colorable with probability $1-o(1)$.
\end{lemma}
\begin{proof}

Let $\delta$ be the value defined by Lemma \ref{lem:BOTlem6} for $K=1$.

By contradiction, suppose that there exists a set $S$ of size $s<\delta n$ such that $H|_{S}$ is not 3-colorable. Assume that $S$ is a minimal set with this property. Suppose that $H|_{S}$ contains a vertex of degree two or less. By the minimality of $S$, there is a 3-coloring of hypergraph $H|_{S-\{v\}}$. However, this coloring  extends to a 3-coloring of $H|_{S}$, by picking a color for $v$ that does not match any of its neighbours. Therefore, it must be that $H|_{S}$ contains exclusively vertices with degree at least 3. Therefore, $H|_{S}$ must contain at least $3n/3=n$ hyperedges, as each hyperedge has 3 vertices, which contradicts Lemma \ref{lem:BOTlem6}. 
\end{proof}

In order to prove that the hypergraph $H$ with probability $1-o(1)$ is not 3-colorable, we bound the probability that a given partition of $V(H)$ is $\varepsilon$-close to a valid coloring of $H$. In particular, we prove the following lemma. 
\begin{lemma}
        Let $H$ be sampled from $\cH$. Given any partition $\{S_{1},S_{2},S_{3}\}$ of $V(H)$, for all $i\in [3]$, we denote by $X_{S_{i}}$ the set of hyperedges contained $S_{i}$. Let $X=X_{S_{1}}+X_{S_{2}}+X_{S_{3}}$, then for every constant $\alpha>0$, the following holds:
        \begin{equation*}
            \mathbb{P}[X<(1/9-\alpha)dn]\leq \exp(-(\alpha-o(1))^{2}dn)
        \end{equation*}
\end{lemma}
\begin{proof}
        Let $I_{1},\dots, I_{dn/3}$ be the sequence of outcomes which reveal the hyperedges of $G$ one by one. For a fixed partition $\{S_{1},S_{2}, S_{3}\}$, the random variable $X$ determines a Doob martingale with respect to the filtration $(I_{i})_{i=1}^{dn/3}$. It follows for all $1<j\leq \frac{d  n}{3}$ that
        \begin{equation*}
            |\mathbb{E}[X|I_{1},\dots, I_{j}]-\mathbb{E}[X|I_{1},\dots, I_{j-1}]|\leq 1.
        \end{equation*}
        Note that, by convexity, $\mathbb{E}[X]\geq\frac{dn(n-3)(n-6)}{9(n-1)(n-2)}$ as this value is minimised when $|S_{1}|=|S_{2}|=|S_{3}|=\frac{n}{3}$ and the probability that each hyperedge is in any given $S_{i}$ is $\frac{n/3 -1}{n-1}  \frac{n/3 -2}{n-2}$. Azuma's identity says that 
        \begin{equation*}
            \mathbb{P}\left[X<\left(\frac{(n-3)(n-6)}{9(n-1)(n-2)}-\beta\right)dn\right]\leq e^{-\beta^{2}dn}.
        \end{equation*}
        The statement follows for $\beta=\alpha-\frac{1}{9}+\frac{(n-3)(n-6)}{9(n-1)(n-2)}=\alpha-o(1)$.
    \end{proof}
Further, we can prove the following. 
  \begin{lemma}
        For any constant $\alpha>0$, there exists a constant $d$ such that with probability $1-o(1)$ any 3-coloring of $\overline{H}$ has at least $(1/9-\alpha)dn$ violating hyperedges.
    \end{lemma}
\begin{proof}
        The number of ways to partition $V(G)$ into $3$ sets is $3^{n}$. Therefore if $d>\frac{ln(3)}{\alpha^{2}}$, then we have that the graph is $1/9 - \alpha$ far from being 3-colorable with high probability. We let $M_{u,v,w}$ be the event that $(u,v,w)$ is a hyperedge with multiplicity at least $2$. $\mathbb{P}[M_{u,v,w}=1]=O(d^{2}/n^{4})$ by a union bound over each $C_{i}$. By Markov, the probability that there are $d  \log(n)$ variables $M_{u,v,w}=1$, is $o(1)$. Every edge has multiplicity at most $d$ so that means that $|E(G)-E(\overline{G})|\leq d^{3}\log(n)=o(n)$. Therefore the conclusion holds.
    \end{proof}
\section{Sketch of the Proof of Theorem \ref{theo: Hardness-k-col}}
\label{app:kcol}
We can now briefly discuss how to extend the result in Theorem \ref{theo: Hardness-3-col} to the setting where the hypergraphs involved are $k$-uniform for a constant $k\geq 4$, i.e., we explain how to prove Theorem \ref{theo: Hardness-k-col}. The starting point for proving Theorem \ref{theo: Hardness-k-col} is Theorem \ref{theo: BOT-SAT} from \cite{BOT02}. The proof strategy is the same as before. Yet, minor modifications of the gadgets are needed that, for the sake of completeness, we describe them below.  The vertex set of $H:=\rho_{k\text{-col.}}(\mathcal{C})$ is structured as before, the difference being that we introduce $k$ distinct families of auxiliary vertices $(T,F, D_1,...D_{k-2})$, each of them corresponding to a distinct color. The gadgets have the same role as before and are structured as follows:
 \begin{itemize}
     \item \emph{\textbf{Equality gadgets}} ensure that the two distinct vertices $u$ and $v$ are assigned the same color within any valid hypergraph $k$-coloring. They consist of vertices $u,v$ and $k^2 -k-1$ auxiliary vertices, and they have a hyperedge for every set of $k$ vertices that doesn't include both $u$ and $v$. 
     \item \emph{\textbf{Inequality gadgets}} ensure that two vertices $u$ and $v$ are assigned a distinct color within any valid hypergraph $k$-coloring. For the two vertices $u,v$, there are $k-2$ auxiliary vertices $u_{1},\ldots,u_{k-2},v_{1},\ldots,v_{k-2}$ among which we insert equality gadgets. Lastly, hyperedges $\{u,u_{1},\ldots,u_{k-1},v\}$ and $\{u,v_{1},\ldots,v_{k-2},v\}$ are added.
     \item \emph{\textbf{Not dummy gadgets}} ensure that a vertex $u$ is assigned color \textsc{True} or \textsc{False}. These gadgets consist of $u$, and $k-1$ vertices of $D_{\text{i.}}$ for all $i\in [k-2]$. 
     \item \emph{\textbf{Clause gadgets}} ensure that any $k$-tuple of literal vertices contained in a clause $C$ is not monochromatically colored with color \textsc{False}. We replicate the structure of clause gadgets for 3-colorability and add to them $k-1$ vertices from each of the dummy classes $D_2,...D_{k-2}$, so that each auxiliary vertex shares a hyperedge with each $(k_1)$-tuple of dummy vertices.
 \end{itemize}

\section{Sketch of the Proof of Theorems \ref{theo: Hardness-k-part} and  \ref{theo: testability-k-par}}\label{app:kpar}
We can now extend the result stated in Theorem \ref{theo: Hardness-3-part} to the setting where the hypergraphs are $k$-uniform for $k\geq 4$, i.e., we can explain how to prove Theorem \ref{theo: Hardness-k-part}. The starting point for proving this result is Theorem \ref{theo: BOT-Col} from \cite{BOT02}. The proof strategy is the same as before, i.e., we exhibit a local and gap-preserving reduction $\rho_{k\text{-par.}}$. Minor modifications are needed, for sake of completeness, we briefly describe them.  

Given a graph $G = ([n], E)$ of maximum bounded degree $d_{3\text{-col.}}$, we construct a $k$-uniform hypergraph $H:=\rho_{\text{3-par.}}(G)=([n+(k-2)d_{3\text{-col.}}n], E')$, as follows: for every edge $(u,v) \in E$, we introduce a hyperedge $\{u,v, n+d_{3\text{-col.}}(j-1)+1, \dots, n+d_{3\text{-col.}}(j-1)+k-2\}$ in $E'$ for $(u,v)\in E$ being the $j$th edge incident at $u$ for $u<v$. Here, for every $i>n$ that does not belong to the vertex set of $G$, each such vertex is contained in at most one unique hyperedge.

Extending the result of Theorem \ref{theo: Hardness-3-part} to $k$-uniform hypergraphs for $k\geq 4$ is straightforward. Indeed, the proof of Theorem \ref{theo: testability-k-par} is the same as the one of Theorem \ref{theo: Hardness-3-part}. 
\end{document}